\documentclass[a4paper,12pt]{article}%

\usepackage[T1,T5]{fontenc}
\usepackage{geometry}
\usepackage{amsmath}
\usepackage{amsfonts}
\usepackage{amssymb}
\usepackage{graphicx}
\usepackage{indentfirst}
\usepackage{bbold}
\usepackage[small,bf]{caption}
\usepackage{slashed}
\usepackage{xcolor}
\usepackage[normalem]{ulem}

\usepackage[normalem]{ulem}
\newcommand\redsout{\bgroup\markoverwith{\textcolor{red}{\rule[0.4ex]{3pt}{0.7pt}}}\ULon}

\providecommand{\U}[1]{\protect\rule{.1in}{.1in}}

\DeclareMathOperator{\Tr}{Tr}
\newcommand{\p}{\partial}

\newcommand{\e}{\ensuremath{\mathrm{e}}}

\usepackage{color}
\definecolor{darkgreen}{rgb}{0,0.35,0}
\definecolor{Rood}{rgb}{1, 0, 0}

\begin{document}

\date{}
\title{\textbf{Lorentz symmetry breaking  in a Yang-Mills theory within the Gribov
restriction}}
\author{
\textbf{D.~R.~Granado}$^{a}$\thanks{diegorochagranado@duytan.edu.vn}\,\,,
\textbf{I.~F.~Justo}$^{b,\,c}$\thanks{igorfjusto@gmail.com}\,\,,
\textbf{A. Yu. Petrov}$^{d}$\thanks{petrov@fisica.ufpb.br}\\[2mm]
{\small \textnormal{$^{a}$  \it Institute of Research and Development}} \\ 
{\small \textnormal{ \it Duy T{a}n University, 3 Quang Trung, D{a} Nang, Vietnam}\normalsize} \\ 
{\small \textnormal{$^{b}$ \it Instituto de Ciencias F\'{i}sicas y Matem\'{a}ticas,}} \\ 
{\small \textnormal{\phantom{$^{a}$} \it Universidad Austral de Chile, Valdivia, Chile}} \\
{\small \textnormal{$^{c}$ \it Universidade Federal Fluminense, Instituto de F\'isica,}} \\
{\small \textnormal{\phantom{$^{c}$} \it Av. Litoranea s/n, 24210-346, Niter\'oi, RJ, Brasil}} \\
{\small \textnormal{$^{d}$  \it Departamento de F\'{i}sica, Universidade Federal da Para\'{i}ba,}} \\ 
{\small \textnormal{ \it Caixa Postal 5008, 58051-970, Jo\~{a}o Pessoa, Para\'{i}ba, Brazil}\normalsize} \\ 
}

\maketitle

\begin{abstract}
In this paper we take into account the existence of Gribov copies in a four-dimensional Yang-Mills theory in the Landau gauge with an effective Lorentz symmetry breaking, induced by the Carroll-Field-Jackiw (CFJ) term. The self consistent gap equation of the Gribov parameter is properly derived and, provided observational bounds to the amplitude of the CFJ parameter in the $SU(2)$ gauge theory, a qualitative analysis of the Gribov parameter and of the poles of the propagator is developed.
\end{abstract}

\section{Introduction}

In the last few years we have witnessed an increase in the interest for Lorentz symmetry breaking models within many contexts. This class of models was firstly proposed in the 90's in the context of QED by Carroll, Field and Jackiw (CFJ) \cite{Carroll:1989vb}. For the first time, the authors introduced a consistent Lorentz-breaking extension of {a} known field theory model by defining a privileged space-time direction by means of a constant axial vector $a_{\mu}$. Extensions of the standard model considering effective Lorentz symmetry breaking were subsequently proposed \cite{Colladay:1996iz,Colladay:1998fq}, and many nontrivial issues related with these theories have been discussed. For instance, we can emphasize an unusual wave propagation including birefringence and rotation of the polarization plane of electromagnetic field in a vacuum \cite{Jackiw:2001dj,Guralnik:2001ax}, which has been shown to take place in various Lorentz-breaking extensions of QED \cite{Myers:2003fd,Casana:2009xs}, ambiguity of quantum corrections \cite{Jackiw:1999qq} and perturbative generating new Lorentz-breaking terms \cite{Carroll:1989vb}. Many possible impacts of Lorentz symmetry breaking have been measured experimentally in different cases \cite{Kostelecky:2008ts}. The renormalizability of Lorentz violating extension of QED was discussed in \cite{Santos:2016bqc}.

As mentioned before, the Lorentz symmetry breaking was treated in the context of the QED and, naturally, one can ask about the possibility of a non-Abelian extension of Lorentz-breaking terms. The non-Abelian CFJ term has been generated perturbatively in \cite{Gomes:2007rv,Mariz:2007gf} and some consequences of including this term were discussed in \cite{Santos:2016dcw,Santos:2016uds,Santos:2014lfa}. The first example of such a theory studied in a systematic manner is the four-dimensional Lorentz-breaking Yang-Mills (YM) theory, originally formulated in \cite{Colladay:2006rk}, whose Lagrangian {is the} sum of the standard Yang-Mills Lagrangian and the non-Abelian generalization of the Carroll-Field-Jackiw (CFJ) term.
In that paper, the authors explored the one-loop renormalization of YM-CFJ system. The renormalizability of non-Abelian systems with Lorentz symmetry breaking was explored in \cite{Santos:2016dcw,Santos:2016uds,Santos:2014lfa}.

By construction, the CFJ-like extension of the Yang-Mills theory does not rule out the gauge freedom of the model. Thus, in order to investigate gauge dependent quantities ($e.g.$ $n$-point Green's function), a gauge condition must be chosen. In this work we will focus in the Landau gauge condition, since in this gauge the Faddeev-Popov (FP) operator is manifestly Hermitian. As a consequence, the existence of Gribov copies must be properly addressed, and this will be the main goal of this work.

At the end of the 70's, V. N. Gribov showed  \cite{Gribov:1977wm} that the standard Faddeev-Popov gauge fixing procedure is not enough to unambiguously fix the gauge
freedom of Yang-Mills theories. He showed that, even after imposing the Landau (and Coulomb) gauge, there still remain redundant gauge fields configurations, called Gribov copies, and that the existence of such gauge fixing residual ambiguity is closely related to the existence of zero-modes of the Faddeev-Popov (FP) operator. Subsequently, I.~M.~Singer came out to the conclusion that ``the Gribov ambiguity for the Coulomb gauge will occur in all other gauges''. That is, still in his words, ``no gauge fixing is possible'' \cite{Singer:1978dk}. In practice, it is well known that linear covariant gauges (LCG) are plagued by Gribov copies, and recently a proper BRST-invariant approach was proposed in order to address infinitesimal Gribov copies in this gauge class 
\cite{Capri:2015ixa,Capri:2015nzw,Capri:2016aqq,Capri:2016aif,Capri:2016gut,Capri:2017bfd}. The original procedure proposed by Gribov to remove such remaining gauge copies leads to a drastic modification of the gluon propagator and to a (soft) breaking of the BRST symmetry \cite{Sorella:2009vt,Dudal:2009bf,Capri:2010hb,Dudal:2012sb,Lavrov:2011wb,Lavrov:2012gb} (as mentioned before, a BRST-invariant formulation was recently proposed).
As a consequence, a gauge mass parameter called the Gribov parameter is introduced and the gauge two-point Green's function becomes plagued by complex conjugated poles, which forbids us to construct  K\"allen-Lehmann spectral representation \cite{Sobreiro:2005ec,Vandersickel:2012tz} since it must be an always positive (see \cite{Hayashi:2018giz} for a recent study on the connection between the existence of complex conjugate poles and a positivity violation of the K\"allen-Lehmann representation)\footnote{The gauge dependence of the Gribov problem has been studied in \cite{Lavrov:2011wb,Lavrov:2013boa}.}. Following Osterwalder and Shrader \cite{Osterwalder:1973dx}, the positivity violation in the K\"allen-Lehmann spectral representation prevents one to attach the physical meaning of asymptotic particle to the gauge field propagator. In this sense, Gribov proposed a ``confinement'' interpretation for the gauge field within his framework. 

Gribov's confinement interpretation has attracted considerable attention in the last few years. According to lattice quantum field theory (QFT), the gauge field propagator is infrared (IR) finite ($i.e.$ massive) and is compatible with propagators displaying complex conjugated poles (Stingl-like propagators), \cite{Cucchieri:2007rg,Cucchieri:2009zt,Cucchieri:2010xr}. The agreement between recent lattice simulations and a refined version of Gribov-Zwanziger (RGZ) approach has been sistematically verified, as can be seen $e.g.$ in \cite{Dudal:2008sp,Dudal:2010tf,Cucchieri:2011ig}\footnote{\textit{Cf.} \cite{Dudal:2008sp,Vandersickel:2012tz} for a complete review on RGZ.}. A deeper investigation of geometrical properties of the Gribov issue within lattice framework was developed in \cite{Cucchieri:2012cb,Cucchieri:2013nja}. It is important to mention that Dyson-Schwinger equation (DSE) technique {is also largely applied to the investigation of gauge field propagators, \cite{Alkofer:2000wg,Watson:2001yv,Alkofer:2003jj}. It is also remarkable that since 2004 DSE schemes are pointing to an infrared massive gluon propagator, nowadays} in agreement with lattice data \cite{Aguilar:2004sw,Aguilar:2006gr,Aguilar:2008xm,Strauss:2012dg} and \cite{Maas:2011se}\footnote{\textit{Cf.} \cite{Huber:2018ned} to a recent nice review on nonperturbative properties of Yang-Mills theories within DSE framework.}. Besides that, an alternative approach to Gribov's issue has been recently developed \cite{Tissier:2010ts,Tissier:2011ey,Pelaez:2014mxa}.

In the present work, the original (perturbative) formulation of Gribov will be applied to an $SU(2)$ YM theory with a CFJ term implementing an effective Lorentz symmetry breaking. Although Gribov's formulation does not agree with lattice results in the deep IR regime, it is in qualitative agreement with the refined Gribov-Zwanziger approach (which does fit lattice data, \cite{Dudal:2008sp,Dudal:2010tf,Cucchieri:2011ig}) in what concerns the confinement interpretation of the gauge field. Therefore, in this paper we are only concerned with the qualitative behavior of the gauge field propagator.

This paper is organized as follows. The structure of the paper looks like follows. in the section 2 the problem of the Gribov restriction in non-Abelian gauge theories is briefly reviewed; in the section 3 we address the Gribov problem within the $SU(2)$ YM theory with a CFJ-like Lorentz breaking term, and section 4 is devoted to the conclusion.

\section{A brief review on the functional restriction to the Gribov horizon}
\label{gribovsection}

In the seminal work \cite{Gribov:1977wm}, Gribov realized that the standard gauge fixing
procedure by Faddeev-Popov is not sufficient to remove all equivalent gauge field
{configurations}. His work was aimed to the Landau (and Coulomb) gauge using the Hermiticity of the FP operator in such a gauge. 

In the present section, we will briefly present Gribov's original construction, starting from
the Faddeev-Popov gauge fixing procedure to the Landau gauge and ending up with the expression 
for the gauge field propagator within Gribov's scheme. For a detailed (and considerably didactic) overview on the Gribov issue, including also the Gribov-Zwanziger and the refined Gribov-Zwanziger approaches, we recommend the Report \cite{Vandersickel:2012tz}.

\subsection{Gribov restriction in Yang-Mills theories}
\label{gribov} 

The Euclidean Yang-Mills path integral reads
\begin{equation}
Z_{YM} ~=~ \int DA \; e^{- \frac{1}{4} \int d^{d}x \; F^{a}_{\mu \nu}F^{a}_{\mu\nu}  }
\;,
\label{ympathintegral}
\end{equation}
with $F^{a}_{\mu\nu}$ standing for the field strength tensor,
\begin{equation}
F^{a}_{\mu\nu} ~=~ \partial_{\mu}A^{a}_{\nu} - \partial_{\nu}A^{a}_{\mu} + gf^{abc}A^{b}_{\mu}A^{c}_{\nu}\;. \label{fstr}
\end{equation}
As it is well known, the partition function \eqref{ympathintegral} is plagued with gauge redundancy. The Faddeev-Popov gauge fixing procedure will be used in order to impose the Landau gauge condition, so that the gauge fixed partition function becomes
\begin{equation}
Z_{FP} ~=~\int DA DcD\bar{c} \; e^{-S_{FP}}
\;,
\label{fppathintegral}
\end{equation}
with
\begin{equation}
S_{FP} ~=~ 
\frac{1}{4} \int d^{d}x \; 
F^{a}_{\mu \nu}F^{a}_{\mu\nu} 
+
\int d^{d}x \;
\left(  b^{a}\partial_{\mu}A^{a}_{\mu}
+\bar{c}^{a} \partial_{\mu}D^{ab}_{\mu}c^{b}  \right)
\,.
\label{gf}
\end{equation}
The fields $({\bar c}^a, c^a)$ are the Faddeev-Popov ghosts; $b^a$ accounts for the Lagrange multiplier implementing the Landau gauge condition,
\begin{eqnarray}
\partial_{\mu}A_{\mu} ~=~ 0
\,;
\label{lgc}
\end{eqnarray}
and $D^{ab}_\mu =( \delta^{ab}\partial_\mu + g f^{acb}A^{c}_{\mu})$ is the covariant derivative in the adjoint representation of $SU(N)$. The gauge fixing contribution to equation \eqref{gf} is introduced in a BRST invariant manner, with $\{ \bar{c}^{a}, b^{a} \}$ being a BRST doublet.

Gribov's great contribution consisted in showing that, even after imposing the gauge condition, the partition function \eqref{fppathintegral} is still plagued by the presence of some physically equivalent gauge field configurations. In other words, he showed that it is possible to find out two gauge field configurations satisfying \eqref{lgc}, $\p_{\mu}A_{\mu} = \p_{\mu}A^{\prime}_{\mu} = 0$, but which are also (infinitesimally) gauge equivalent configurations: $A^{\prime}_{\mu}(x) = A_{\mu}(x) - D_{\mu}\omega(x)$, with $\omega(x)$ standing for the field that implements the infinitesimal gauge transformation.

In order to get rid of these remaining ambiguities, Gribov proposed to restrict the path
integral of the gauge field to the region where the FP operator ${\cal M}^{ab}$ is positive
definite. This region is the so called ``first Gribov region'', defined as
\begin{align}
\Omega ~=~  \{ A^a_{\mu}\,; ~ 
\partial_\mu A^a_{\mu}=0 \,; ~
{\cal M}^{ab} = -\partial_{\mu}(\partial_{\mu}
\delta^{ab} -g f^{abc}A^{c}_{\mu})\; >0 \} \;. 
\label{gr}
\end{align} 
Such region is supposed to be free of infinitesimal copies, since we are considering copies associated with infinitesimal gauge transformations of the gauge field. The region formally free of all possible Gribov copies (infinitesimal and finite ones) is called the Fundamental Modular Region (FMR). So far, there is no viable approach to implement the restriction of the path integral to the FMR, to our knowledge  \cite{Zwanziger:1993dh,Dudal:2008sp,Cucchieri:1997ns}.

From the definition \eqref{gr}, it is possible to see that a gauge field configuration with
large enough amplitude may correspond to negative eigenvalues of the FP operator,  {\textit{i.e.},} it is
outside $\Omega$. Therefore, Gribov's restriction means that only gauge field configurations
with small enough amplitude, so that $\mathcal{M}^{ab} > 0$, must be accounted in the path integral.

Since the FP operator is closely related to the inverse of the ghost field propagator, one may realize that the Gribov restriction condition amounts to restrictions on the ghost-anti-ghost two-point function. That is, Gribov proposed to investigate, order by order in loop corrections, whether gauge field configuration is associated to divergencies of the ghost-anti-ghost propagator, which means that one should perturbatively access the influence of $A_{\mu}(x)$ on the poles of the ghosts field propagator. In order to do that, the ghost two-point function is computed up to one-loop order as a functional of the gauge field, \textit{i.e.} the gauge field is treated as a classical field at this moment,
\begin{eqnarray}
\mathcal{G}(k,A) &=&
\frac{\delta^{ab}}{N^2-1}\frac{1}{k^2}\left(\delta^{ab}+\sigma^{ab}(k,A)  \right) 
\;,
\label{ghostformfactor1}
\end{eqnarray}
where $\sigma(k,A)$ stands for the trace of the ghost form factor. The idea is to consider small enough amplitudes of the gauge field $A_{\mu}(x)$ so that the following approximation becomes reasonable,
\begin{eqnarray}
\mathcal{G}(k,A)
~\approx~
\frac{1}{k^2}\frac{1}{\left(1-\sigma(k,A)\right)} 
\,.
\label{ghostformfactor}
\end{eqnarray}
Then, Gribov's proposal is that one should consider only gauge field configurations corresponding to
\begin{equation}
\sigma(k,A) ~<~ 1
\;,
\label{npcondition}
\end{equation}
in order to ensure the positivity of the FP operator. Notice that the condition \eqref{npcondition} imposes that the only possible pole of the ghost-anti-ghost propagator is at $k^{2} = 0$, and therefore it is called the ``no-pole condition''. The pole structure of the ghost-anti-ghost propagator has been investigated during a long time on the lattice, and recent results are pointing to a ``non-enhanced'' propagator, \cite{Cucchieri:2007md,Cucchieri:2007rg,Cucchieri:2008fc,Bogolubsky:2009qb}, in disagreement with Gribov's original prediction, but consistently with the refined version of the Gribov-Zwanziger formulation, \cite{Dudal:2008sp}.

The idea to avoid the residual Gribov ambiguity after fixing the Landau gauge is to restrict the path integral to a domain where the gauge field configuration satisfies the no-pole condition \eqref{ghostformfactor}  (\textit{i.e.} the First Gribov region $\Omega$). Such restriction is implemented by means of the introducton of the Heaviside step function, namely,
\begin{eqnarray}
Z_{G} &=&
\int_\Omega DA DcD\bar{c} \; e^{-S_{FP}}
\nonumber\\
&=&\int DA DcD\bar{c} \;  \theta(1 - \sigma(k,A)) e^{-S_{FP}}
\,.
\label{gribovpathintegral}
\end{eqnarray}

The ghost-anti-ghost two-point function computed in the presence of external gauge field reads,
up to the first order in the quantum fields, looks like
\begin{equation}
\mathcal{G}(k,A)=\frac{1}{k^2}\left(1+\frac{k_\mu k_\nu}{k^2}\frac{Ng^2}{Vd(N^2-1)}\int
\frac{d^dp}{(2\pi)^4}\frac{A_\mu^a(p)A_\nu^a(-p)}{(k-p)^2}\right) \,.
\label{ghostformfactor1pi}
\end{equation}
One should notice that the ghost form factor decreases as $k^{-2}$, so that the no-pole
condition may be taken in the limit $k \to 0$. Taking the limit $k\to0$, the ghost form factor reads\footnote{
See \cite{Vandersickel:2012tz} for a detailed computation of
\eqref{ghostformfactor1pi} and for further discussions.
} 
\begin{equation}
\sigma(0,A) ~=~ 
\frac{Ng^2}{dV(N^2-1)}\int
\frac{d^dp}{(2\pi)^d}\frac{A_\mu^a(p)A_\mu^a(-p)}{p^2} 
\, .
\label{ghostformfactor2}
\end{equation}

Considering the integral representation of the Heaviside step function, the Gribov partition function becomes
\begin{eqnarray}
Z_{G} ~=~
\int 
\int^{\infty+i\epsilon}_{-\infty+i\epsilon}\frac{d\beta}{2\pi i\beta}
DA DcD\bar{c} 
\; e^{\beta(1-\sigma(0,A))}e^{-S_{FP}}.
\label{gribovpathintegral1}
\end{eqnarray}
The integral over $\beta$ will be performed by means of the saddle-point approximation, which becomes exact in the thermodynamic limit where the microcanonical ensemble becomes equivalent to the canonical Boltzmann ensemble \cite{Dudal:2009xh}.

From equation \eqref{gribovpathintegral1}, one may read off the Gribov action (in general $d$-dimension),
\begin{eqnarray}
S_{G} ~=~ 
S_{FP}
+
\frac{\beta Ng^2}{dV(N^2-1)}
\int d^{d}x \; 
A_\mu^a(x) \left[\p^{2}\right]^{-1} A_\mu^a(x)
-\beta
\,.
\label{wrhglwieh}
\end{eqnarray}
Notice that the Gribov action is manifestly non-local, which is not actually a problem since it can be converted to a  local form by introducing auxiliary fields. Such non-locality may indicate that the Gribov action \eqref{wrhglwieh} takes into account IR effects of the non-Abelian model. An interesting consequence of this non-locality will be discussed in the following subsection.

\subsection{The gap equation and the gauge field propagator}
\label{glounpropagatorsec}

One important point of the Gribov approach is that the mass parameter $\beta$, introduced for the implementation of the Heaviside step function, is not a free parameter of the theory. Otherwise, the no-pole condition \eqref{npcondition} implies a consistent determination of $\beta$. That is, considering the action \eqref{wrhglwieh} for a free parameter 
$\beta$ the no-pole condition is not immediately satisfied. In order to impose this condition, one must integrate over the $\beta$ integral in the partition function \eqref{gribovpathintegral1}.

At the tree-level in perturbation theory, the partition function \eqref{gribovpathintegral1} can be written as
\begin{align}
Z_{G} ~=~ 
\int \!\!\!
\int^{\infty+i\epsilon}_{-\infty+i\epsilon}
\frac{d\beta}{2\pi i\beta}     DA DcD\bar{c}      \,
\exp  \left\{
-
\int \frac{d^dp}{(2\pi)^d)}
\left[
 \frac12
A^{a}_{\mu}(p)
Q^{ab}_{\mu\nu}
A^{b}_{\nu}(-p)
+
\bar{c}^{a}(p)
P^{ab}
c^{b}(-p)
\right]
-
\beta
\right\}
\,,
\end{align}
with
\begin{equation}
Q_{\mu\nu}^{ab} ~=~ 
\delta^{ab}
\left[
\left(\frac{2N \beta g^2}{Vd(N^2-1)}\frac{1}{p^2}+p^2\right)\delta_{\mu\nu}+\left(\frac{1}{\alpha}-1\right)p_\mu p_\nu
\right]
\label{gluonoperator}
\end{equation}
and
\begin{align}
P^{ab} ~=~
- \delta^{ab} \p_{\mu}\p_{\mu} 
\,.
\end{align}
Integrating out the fields, one ends up with
\begin{eqnarray}
Z_{G} ~=~
\int^{\infty+i\epsilon}_{-\infty+i\epsilon}
\frac{d\beta}{2 i \pi}
\left[  \det  Q^{ab}_{\mu\nu} \right]^{-1/2}
\left[  \det P^{ab} \right]
\e^{\beta - \ln\beta}
\,.
\end{eqnarray}
Making use of the following identify,
\begin{equation}
\left[
\det Q_{\mu\nu}^{ab}
\right]^{-1/2}=e^{-\frac{1}{2}\det\ln Q_{\mu\nu}^{ab}} ~=~
\e^{-\frac{1}{2} \Tr\ln Q_{\mu\nu}^{ab}}
\label{weighwg}
\end{equation}
one rewrites the path integral as
\begin{align}
\e^{-V{\cal E}_{v}} ~=~ Z_{G} ~=~ 
\int^{\infty+i\epsilon}_{-\infty+i\epsilon}
\frac{d\beta}{2 i \pi}
\e^{-f(\beta)}
\,,
\end{align}
with
\begin{align}
f(\beta) &=
\beta-\ln\beta-\frac{d-1}{d}(N^2-1)V\int \frac{d^dp}{(2\pi)^d}\ln\left(p^2+\frac{\beta Ng^2}{N^2-1}\frac{2}{dV}\frac{1}{p^2}\right)
\,.
\end{align}
As mentioned before, the $\beta$ integral can be performed by means of the saddle-point approximation, which becomes exact in the thermodynamic limit. Namely,
\begin{equation}
Z_{G} ~\approx~ e^{-f(\beta^{\ast})},
\end{equation}
where $\beta^{\ast}$ satisfies the saddle-point equation,
\begin{equation}
\frac{\p {\cal E}_{v}}{\p \beta} \Bigg \vert_{\beta = \beta^{\ast}} ~=~ 0
\,,
\end{equation}
leading us to the called gap equation,\footnote{In the thermodynamic limit the term
$\ln\beta$ can be disregarded, since $\beta$ is proportional to the volume $V$.}
\begin{equation}
1 ~=~ \frac{d-1}{d}Ng^2
\int \frac{d^dp}{(2\pi)^d)}\frac{1}{p^4+\gamma^{\ast \, 4}}
\,.
\label{gapequation}
\end{equation}
Just to simplify the notation, we defined $\gamma^4=\frac{2g^2\beta N}{dV(N^2-1)}$ in the above gap equation and from now on we will use the $\gamma^{4}$ notation. The gap equation \eqref{gapequation} is a self consistent condition that the Gribov mass parameter must satisfy. It means that, after imposing the no-pole condition on the ghost two-point function (given at equation \eqref{ghostformfactor}), the Gribov parameter is dynamically fixed, in the sense that the gap equation must be computed order-by-order in loop corrections.

Now, after fixing the Gribov parameter, which makes consistent the model, we are able to analyze the gauge field propagator. Namely, finding the inverse of \eqref{gluonoperator} and taking the Landau limit $\alpha \to 0$ at the very end of the computation, we get the result
\begin{equation}
\langle A_\mu^a(k)A_\nu^b(-k)\rangle ~=~ 
\delta^{ab}\frac{k^2}{k^4+\gamma^{\ast \, 4}}
\left(\delta_{\mu\nu}-\frac{k_\mu k_\nu}{k^2}\right)
\,,
\end{equation}
which can be rewritten as
\begin{equation}
\langle A_\mu^a(k)A_\nu^b(-k)\rangle ~=~ 
\delta^{ab}
\frac{1}{2}\left(\frac{1}{k^2+i\gamma^{\ast \, 2}}+\frac{1}{k^2-i\gamma^{\ast \, 2}}\right)
\left(\delta_{\mu\nu}-\frac{k_\mu k_\nu}{k^2}\right)
 \,.
\label{gribovpropagator}
\end{equation}
From \eqref{gribovpropagator}, it is clear that after Gribov's restriction, the gluon propagator
displays complex conjugate poles, which prevents to assign an asymptotic single particle interpretation to it, in
the sense that its K\"all\'en-Lehmann representation is not always positive \cite{Sorella:2010it}.

\section{The Lorentz symmetry breaking: 4D Carroll-Field-Jackiw model}

As mentioned at the introduction, the seminal work of Carroll, Field and Jackiw  \cite{Carroll:1989vb} was the first theoretical model with a Lorentz symmetry breaking term. 
The additive CFJ term
represents itself as a
natural four-dimensional generalization of the well-known three-dimensional Chern-Simons term,
displaying its main feature, that is, the gauge invariance. 
The CFJ Lagrangian under the gauge transformations varies by a total
derivative. Afterwards, many issues related with the CFJ extended QED have been studied, for instance, dispersion relations (discussed for the first time already in \cite{Carroll:1989vb}), 
generation of the CFJ term as a quantum correction in a special extension of QED, by means of 
spinors coupled to a constant axial vector $b_{\mu}$ \cite{Jackiw:1999yp} and the ambiguity of
this term caused by the fact that the CFJ term arises from a formally logarithmically
divergent, but actually finite contribution displaying the undetermined form $\infty-\infty$
(see the discussion in \cite{Jackiw:1999qq}). Further, the CFJ term has been calculated within
different regularization schemes, and various (always finite) results for it have been obtained (for
details of calculation of this term within different approaches see f.e. \cite{Mariz:2005jh} and
references therein). 

At the same time, it is well known that the three-dimensional Chern-Simons term admits a
non-Abelian generalization. Therefore, it is natural to suggest that the non-Abelian
generalization of the CFJ term can be achieved. Originally, the non-Abelian CFJ term has been
introduced in \cite{Colladay:2006rk} and \cite{Gomes:2007rv}, where it has been shown to be 
{a} quantum {correction} 
(a straightforward non-Abelian generalization of the results of \cite{Mariz:2005jh}). Moreover,  in
\cite{Gomes:2007rv} the generation of the non-Abelian CFJ term for the finite temperature case
has been performed. In \cite{Santos:2016uds} the authors studied the renormalizability  of the Yang-Mills theory in the presence of the non-Abelian CFJ term.

The presence of the non-Abelian CFJ term (and a consequent nonlinearity of the equations of
motion together with their dependence on Lorentz-breaking parameters) clearly establishes the
question about its possible impact  for the quantization procedure of the gauge field. Namely, 
our goal in this paper consists of studying the role played by the CFJ term in the Gribov quantization prescription of gauge fields
within the Landau gauge. We will show that the Lorentz-breaking nature of the CFJ term implies in a strong modification of the two-point Green's function of the gauge field, whose behavior will depend strongly on the CFJ coupling constant. Besides the similarity with the non-Abelian Chern-Simons term, the CFJ term is not a topological one. In the paper where the term was proposed, Carroll, Field and Jackiw \cite{Carroll:1989vb} showed that this term contributes to the energy-momentum tensor (EMT). The CFJ contribution turns the EMT into a quantity without Lorentz invariance. Thus, the energy density of the system is affected by this term. In our scenario this non-topological feature of the CFJ term will be explored in the gap equation of the Gribov parameter in the section \ref{gapeq}.

%
%
The $4D$ Euclidean non-Abelian Carroll-Field-Jackiw (CFJ) term reads 
\footnote{As we mentioned before, the former term can be obtained by evaluating the fermion
determinant \cite{Gomes:2007rv}.}
\begin{equation}
S_{CFJ} ~=~    
-\frac{i}{2}	
\int d^4x  \;
\xi a_\rho\epsilon_{\rho\mu\nu\lambda}\left(A_\nu^a\partial_\lambda A_\mu^a-\frac{2}{3}gA_\mu^aA_\nu^bA_\lambda^cf^{abc}\right).
\label{ymcfjaction}
\end{equation}
where $\xi$ plays the role of the CFJ coupling constant and $a_\rho$ is the constant vector that determines a specific direction in space-time, which leads to the breaking of the Lorentz symmetry. Here we adopt $a_\rho$ as a dimensionless vector and $\xi$ the CFJ parameter with mass dimension 1.

The issue of the gauge fixing ambiguities, \textit{aka} the Gribov problem, in the presence of background field has been the subject of recent investigations, \cite{Dudal:2017jfw,Kroff:2018ncl}, and it remains not fully understood. As we have chosen the Landau gauge (and not the Landau-de Witt one as in the aforementioned works  \cite{Dudal:2017jfw,Kroff:2018ncl}), the ghost field dynamics remains, a priori, unchanged regarding the standard scenario without the background field. Therefore, in this scenario, it can be seen that the Lorentz symmetry breaking does not influence the quantization procedure of the gauge
field. As the path integral is still plagued by the gauge redundancy (\textit{i.e.} 
gauge and Gribov copies), the FP and Gribov procedures must be applied. Furthermore, reinforcing our previous comment, the presence of the CFJ term in the action is completely harmless to the dynamic of the ghost sector and, in this way, the ghost form factor \eqref{ghostformfactor2} remains the same. Therefore, the total action (after imposing the Gribov restriction) reads,
\begin{eqnarray}
S ~=~ S_{G} + S_{CFJ}
\,,
\end{eqnarray}
with $S_{G}$ given by \eqref{wrhglwieh}.

\subsection{The gap equation}
\label{gapeq}

In this section we analyze the contribution of the CFJ term to the gap equation. The energy momentum tensor indeed is affected, since the CFJ term is not a topological invariant. Therefore, the corresponding gap equation can (and will) be changed by the CFJ Lorentz symmetry breaking term. That is the goal of this subsection. 


In order to compute the gap equation, either at the tree level or within loop corrections, the quantum fields must be integrated out in the  partition function. Since no loop correction will be considered here, only quadratic terms in the quantum fields will be considered (as well as constant terms). 
By following the step by step the procedure described in  the section \ref{gribovsection}, we have
\begin{equation}
S ~=~ \int \frac{d^4p}{(2\pi)^4}
\left\{
-\frac{1}{2}\tilde{A}_\mu^a(p)Q_{\mu\nu}^{ab}\tilde{A}_\nu^b(-p)
+ 
\bar{c}^{a}(p)    P^{ab}  c^{b}(-p)
- \beta + \ln \beta
\right\}
\end{equation}
 where 
 \begin{equation}
 Q_{\mu\nu}^{ab} ~=~ \delta^{ab}\left(\frac{p^4+ \gamma^{4} }{p^2}\delta_{\mu\nu}+\left(\frac{1}{\alpha}-1\right)p_\mu p_\nu-\xi a_\rho\epsilon_{\rho\nu\mu\lambda}p_\lambda  \right)
 \label{q}
 \end{equation}
and
\begin{eqnarray}
P^{ab} ~=~ 
- \p_{\mu}\p_{\mu}  \delta^{ab}
\;,
\end{eqnarray}
keeping in mind the definition $\gamma^{4} = \frac{2Ng^2 \beta}{dV(N^2-1)}$.

After integrating out the fields, one can read off the vacuum energy from its definition,
\begin{eqnarray}
\e^{-V\mathcal{E}_{v}} ~=~ 
\left[  \det  Q^{ab}_{\mu\nu} \right]^{-1/2}
\left[  \det \mathcal{P}^{ab} \right]
\e^{\beta - \ln\beta}
\,.
\end{eqnarray}

Using again the identity \eqref{weighwg}, and then taking the thermodynamic limit, we get the result
\begin{eqnarray}
\mathcal{E}_{v} 
&=&
\ln \beta
- \beta 
+ \frac{1}{2}(N^2-1)
\int \frac{d^{d}p}{(2\pi)^{4}}\; \ln
\left(
\frac{p^{4} + \gamma^{4} }{p^{2}}
\right)
\nonumber \\
&+&
\frac12(N^2-1)(d-2)
\int \frac{d^{d}p}{(2\pi)^{4}}\;
\ln  
\left\{
\left(
\frac{p^{4} + \gamma^{4} }{p^{2}}
\right)^{2} 
+
p^{2} a^{2}\xi^{2}\sin^{2}(\phi)
\right\}
\,.
\label{hwrelgiwhe}
\end{eqnarray}

As mentioned before, the gap equation is a self-consistent condition obtained through the saddle-point approximation condition (which becomes exact in the thermodynamic limit) and it can be obtained by taking the first derivative of equation \eqref{hwrelgiwhe} with respect to $\beta$, computed at the specific value $\beta^*$ which minimizes $\mathcal{E}_{v} $, namely, 
\begin{eqnarray}
&&
\frac{Ng^2}{d}
\int \frac{d^{d}p}{(2\pi)^{d}}\; 
\frac{1}{p^{4} + \gamma^{4} }
+ 
\frac{N(d-2)g^2}{d}
\int \frac{d^{d}p}{(2\pi)^{d}}\;
\frac{p^{4} + \gamma^{4} }{\left( p^{4} + \gamma^{4}  \right)^{2} 
+ p^{6} a^{2}\xi^{2}\sin^{2}(\phi)
}
~=~ 
1
\,.
\label{gpequation}
\end{eqnarray}
From the last term of equation \eqref{gpequation}, one can clearly see that the system is not isotropic anymore, due to the presence of the factor $a^{2}\xi^{2}\sin^{2}(\phi)$, and that the integral must be carefully performed in the angular direction $\phi$. This lack of isotropy is a somewhat expected effect in Lorentz symmetry breaking scenarios, even though we have not chosen any preferred direction. Since such anisotropy does not spoil the power counting divergence of the integral, dimensional regularization can be applied in order to identify the finite contribution of equation \eqref{gpequation}. Thus, we first integrate the radial (momentum) and polar angular directions, while the azimuthal $\phi$-direction can be the last one to be integrated.

In order to simplify computation, let us decompose both integrals in primitive fractions, so that equation \eqref{gpequation} can be rewritten as
\begin{eqnarray}
\frac{Ng^2}{d}
\int \frac{d^{d}p}{(2\pi)^{d}} 
\sum_{i=1}^{2}\frac{\alpha_i}{p^2+m_i^2}
+
\frac{N(d-2)g^2}{d}
\int \frac{d^{d}p}{(2\pi)^{d}} 
\sum_{j=3}^{6}\frac{\alpha_j}{p^2+m_j^2}
\;,
\label{gkrhgigh}
\end{eqnarray}
The poles $m_{i}^{2}$ are functions of $\beta$, $g$ and $a^{2}\xi^{2}\sin^{2}(\phi)$, as well as are the factors $\alpha_i$. Due to the different numerical factors of each integral, we keep separate the factors from the first and second integrals of equation \eqref{gkrhgigh}. Particularly, we define $m_{1}^{2}$, $m_{2}^{2}$ and $\alpha_1$, $\alpha_2$ as
\begin{eqnarray}
m_{1}^{2} ~=~ - m_{2}^{2} ~=~ -i \gamma^{2}
\qquad \text{and} \qquad
\alpha_{1} ~=~ -\alpha_{2} ~=~ 	\frac{1}{4i \gamma^{2}}   \frac{}{}
\;,
\end{eqnarray}
with $\gamma^4 = \frac{2Ng^2\beta }{d(N^2-1)}$.

The second integral will be decomposed in terms of the remaining four pair of parameters $(m_{j}^{2},\ \alpha_{j})$ (for $j = 3,\ 4,\ 5,\ 6$). The four $- m_{j}^{2}$ parameters account for the four roots of the polynomial expression in the denominator of the second integral of equation \eqref{gpequation} and the four $\alpha_j$ stands for the numerical factors. Namely, the factors $\alpha_j$ can be written in terms of the poles $m_j^2$ as
\begin{eqnarray}
\alpha_3&=&\frac{- \gamma^{4}-m_3^4}{(m_3^2-m_4^2)(m_3^2-m_5^2)(m_3^2-m_6^2)} 
\nonumber 
\\
\alpha_4&=&\frac{\gamma^{4} +m_4^4}{(m_3^2-m_4^2)(m_4^2-m_5^2)(m_4^2-m_6^2)} 
\nonumber
\\
\alpha_5&=&\frac{\gamma^{4}+m_5^4}{(m_3^2-m_5^2)(-m_4^2+m_5^2)(m_5^2-m_6^2)} 
\nonumber
\\
\alpha_6&=&\frac{\gamma^{4}+m_6^4}{(m_4^2-m_6^2)(m_6^2-m_3^2)(m_6^2-m_5^2)} 
\;,
\end{eqnarray}
The explicit expression of $\alpha_j (\beta,g,\xi,\phi)$ and $m^{2}_j (\beta,g,\xi,\phi)$ are omitted here, since they are not particularly useful as well as are considerably huge in size.

At this point dimensional regularization can be applied, by taking the radial ($p^{2}$) integral in $d = 4 - \varepsilon$ dimensions, with $\varepsilon \to 0$. Subsequently, the angular integrals can be safely performed. First, note that the gap equation \eqref{gpequation} can be written as
\begin{equation}
Ng^2I^d_1+Ng^2I^d_2 ~=~ 1
\end{equation}
where the first integral $I^d_1$, using dimensional regularization ($d=4-\epsilon$), reads
\begin{eqnarray}
Ng^2I^d_1&=&
\frac{Ng^2}{d}  \int \frac{d^{d}p}{(2\pi)^{d}}\; 
\frac{1}{p^{4} +  \gamma^{2}}
\nonumber\\
&=&
\frac{ Ng^{2} }{4(4\pi)^{2}}
\left(  \frac{2}{\varepsilon} + \frac{3}{2}   -  \ln \frac{\gamma^{2}}{\bar{\mu}^{2}}   + {\cal O}(\varepsilon)   \right)
\label{egrnoig}
\end{eqnarray}
where $\bar{\mu}^2$ is the $\overline{\text{MS}}$  renormalization scale. From the gap equation \eqref{gpequation}, it can be {seen} that the integral $I^d_2$ is the one whose integrand depends on the azimuthal angle. 
Therefore, the dimensional regularization must be performed carefully, leaving the azimuthal integration untouched. Namely,
\begin{eqnarray}
Ng^{2}I^d_2 &=& 
Ng^{2} \frac{(d-2)}{d}   \int \frac{d^{d}p}{(2\pi)^{d}}\;
\frac{\left(p^{4} + \gamma^4\right)}{\left(p^{4} + \gamma^4\right)^{2}
+
p^{6} a^{2}\xi^2 \sin^{2}(\phi)}
\nonumber \\
&=&
\frac{ Ng^2 \pi }{ 32 \pi^{4} } \sum_{i=1}^{4}   \int_0^{2\pi}    ~
m_{i}^{2}\alpha_{i}
\left(  \frac{2}{\varepsilon} + \frac12   - \ln \frac{m_{i}^{2}}{\bar{\mu}^{2}}   \right)
~ d\phi
\label{lerhg}
\end{eqnarray}
Taking the limit $\varepsilon \to 0$ and performing $\overline{\text{MS}}$ renormalization prescription on \eqref{egrnoig} and \eqref{lerhg} we get
\begin{eqnarray}
Ng^2I^d_1&=&
 \frac{ Ng^{2} }{4(4\pi)^{2}}
\left(   \frac{3}{2}   -  \ln \frac{\gamma^{2}}{\bar{\mu}^{2}}   \right)
\,.
\end{eqnarray}
and
\begin{eqnarray}
Ng^{2}I^d_2 &=& 
\frac{ Ng^2 \pi }{ 32 \pi^{4} } \sum_{i=1}^{4}   \int_0^{2\pi}    ~
m_{i}^{2}\alpha_{i}
\left(    \frac12   - \ln \frac{m_{i}^{2}}{\bar{\mu}^{2}}   \right)
~ d\phi
\end{eqnarray}

Finally, the renormalized gap equation, for the specific symmetry group $SU(2)$, reads
\begin{eqnarray}
 \frac{ g^{2} }{32\pi^{2}}
\left(   \frac{3}{2}   -  \ln \frac{\gamma^{2}}{\bar{\mu}^{2}}   \right)
+
\frac{ g^2 \pi }{ 16 \pi^{4} } \sum_{i=1}^{4}   \int_0^{2\pi}    ~
m_{i}^{2}\alpha_{i}
\left(    \frac12   - \ln \frac{m_{i}^{2}}{\bar{\mu}^{2}}   \right)
~ d\phi
~=~ 1
\,.
\end{eqnarray}

In principle, in order to investigate the effects on the Gribov parameter for considering the Carroll-Field-Jackiw  term, in the Landau gauge, the equation \eqref{gpequation} should be solved exactly (for all values of $\phi$). However, the existence of the term $a^{2}\xi^{2}\sin^{2}(\phi)$ in equation \eqref{gpequation}, which is present inside the parameters $\alpha_{j}$ and $m_{j}$ ($j=3,~4,~5,~6$), turns the task of integrating the $\phi$-direction a rather difficult one. Thus, we propose to investigate perturbative contributions to the usual Lorentz invariant $SU(2)$ Gribov parameter. That is, we propose to expand equation \eqref{gpequation} around $\xi = 0$. Such proposal is based on observational bounds of the LSB parameter, $\xi  < 1.1 \times 10^{-7}$GeV, which is considerably smaller than the electroweak energy scale ($\approx 100\text{GeV}$), \cite{Kostelecky:2008ts,Colladay:2018jic}.

From now on we will take $N=2$ and $d=4$. For qualitative analysis, it will be useful to define dimensionless quantities, such as $\bar{\beta}^{\frac12} = \beta^{\frac12}/\overline{\mu}^{2}$ and $\bar{\xi} = \xi/\overline{\mu}$. Thus, after the expansion, keeping terms up to squared powers of $\xi$ and finally integrating the $\phi$-direction, we get 
\begin{eqnarray}
-\frac12 + \frac{32 \pi^{2}}{g^{2}} + \frac{3}{2}\ln \frac{g^{2}\bar{\beta}}{3} 
-
\bar{\xi}^{\, 2}a^{2}\frac{ (3i + 5)  (1 + i\pi - \ln \frac{g^{2}\bar{\beta}}{3}) \sqrt{3} }{32 \sqrt{g^{2}\bar{\beta}}}
~=~ 0
\label{lehgig34}
\end{eqnarray}
Notice that at the Lorentz symmetric limit $\bar\xi \to 0$, the usual Gribov gap equation ($i.e.$ in the Lorentz symmetric Yang-Mills theory) is recovered,
\begin{eqnarray}
\bar{\beta}^{\frac12}_{0} ~=~ \frac{\sqrt{3}}{g} \e^{-\frac{32\pi^{2}}{3g^{2}} + \frac{1}{6} }
\,,
\end{eqnarray}
where $\bar{\beta}_{0}^{\frac12} = \beta_{0}^{\frac12}/\overline{\mu}^{2}$ stands for the ``standard'' dimensionless Gribov parameter in the Lorentz symmetric $SU(2)$ scenario. 

Since in the gap equation \eqref{gpequation} the factor $a^{2}\sin^{2}(\phi)$ does only appear multiplied by $\bar{\xi}^{2}$, and that the integration in $\phi$-direction is performed afterwards the expansion, the first contribution, after considering a small enough parameter $\bar\xi$, is of quadratic order. Thus, only even power terms survive. In order to qualitatively probe the behavior of $\beta^{\frac12}$, we have numerically solved equation \eqref{lehgig34} for random values of $\bar{\xi}$, around the observational bound $\xi < \overline{\mu} \times 10^{-7}$ GeV (considering $\overline{\mu}$ with magnitude of the order of the electroweak energy scale $10^{2}$ GeV), and of $g$. As a result, we could find no significant difference between $\bar{\beta}^{\frac12}$ and $\bar{\beta}^{\frac12}_{0}$. As can be seen from equation \eqref{lehgig34}, the existence of $\bar\xi$ leads to the appearance of an imaginary component of the Gribov parameter, which clearly does not exist in the Lorentz symmetric counterpart $\beta_{0}$. Nonetheless, for the typical order of magnitude of $\bar{\xi}$, such imaginary component appears with power smaller than $10^{-11}$.

Finally, the qualitative behavior of the four poles of the propagator, given by equation \eqref{polynomialequation}, are investigated according to numerical results of equation \eqref{lehgig34}. As expected, the four poles are complex conjugate, for values of $\bar{\xi}$ around $10^{-9}$ and random values of $g$; only for $\bar{\xi}$ with power bigger than $10^{-7}$ a no-zero real component of $\bar\beta$ appears, notwithstanding it is of order $10^{-8}$. Therefore, for the typical order of magnitude of the Lorentz symmetry breaking parameter $\xi$ in the $SU(2)$ gauge group, bounded by observations, \cite{Kostelecky:2008ts,Colladay:2018jic}, the confinement interpretation related to the Gribov type propagator in the Yang-Mills (Lorentz symmetric scenario) scenario can still be completely applied in our case.

\subsection{The gauge propagator}

In order to obtain the propagator, we 
have to
compute the inverse of \eqref{q}, which can be obtained through the following expression,
 \begin{equation}
 Q_{\mu\nu}^{ab}(Q_{\nu\delta}^{bc})^{-1} ~=~ \delta^{ac}\delta_{\mu\delta} \,.
 \label{inversecondition}
 \end{equation}

The Landau gauge is recovered in the limit $\alpha\to0$ and 
the propagator reads
\begin{eqnarray}
\langle A^a_\mu(p) A^b_\nu(-p)\rangle &=& \delta^{ab}{F(p)}
\left[\left(\delta_{\mu\nu}-\frac{p_\mu p_\nu}{p^2}\right)+
\frac{p^6 }{(p^4+\gamma^{\ast \, 4})^{2}}
\left[ \frac{(\xi a\cdot p){p_\mu}}{p^{2}} - \xi a_\mu  \right]
\left[ \frac{(\xi a\cdot p){p_\nu}}{p^{2}} - \xi a_\nu  \right]
\right.
\nonumber\\
&+&
\left.\frac{p^4}{(p^4+\gamma^{\ast \, 4})}a_\beta\epsilon_{\beta\nu\mu\alpha} \frac{\xi p_\alpha}{p^{2}}
\right]  \,.
\label{gribovcspropagator}
\end{eqnarray}
Notice that the overall factor $F(p)$ can be written as
\begin{equation}
F(p)=\frac{(p^4+\gamma^{\ast \, 4})p^2}{(p^4+\gamma^{\ast \, 4})^2+p^6\xi^{2}a^{2}\sin^{2}(\phi)}  \,,
\label{poles}
\end{equation}
with $(a \cdot p) ~=~ \vert a\vert \vert p\vert \cos\phi$.  As it can be verified the propagator \eqref{gribovcspropagator} is transversal to a momentum $p_\mu$, \textit{i.e.}, $p_{\mu}\langle A^a_\mu(p) A^b_\nu(-p)\rangle = 0$. This property is a consequence of the Landau gauge.

It is interesting to notice that the gauge field propagator \eqref{gribovcspropagator} shares some similarities with the gauge field propagator obtained in \cite{Canfora:2013zza}. The authors of \cite{Canfora:2013zza} investigate the poles of the gauge field propagator in the $3$D Yang-Mills-Chern-Simons theory, in the Landau gauge, while taking into account the existence of the Gribov ambiguities. Both propagators have the same overall factor \eqref{poles}, thus presenting the same pole structure 
\eqref{polynomialequation}, and display a parity symmetry breaking structure, and then the analogies cease. It must be clear that the CS term in the model studied in \cite{Canfora:2013zza} is topological and, therefore, does not influence the vacuum energy of the system. Thus, the gap equation found in \cite{Canfora:2013zza} is exactly the same as \eqref{gapequation} with $d=3$. Contrarily to this, we are dealing with a $4$-dimensional YM $+$ CFJ model, which is \textit{not} a topological system. Therefore, as we could see, our gap equation becomes strongly modified, so that the Gribov parameter computed here, formally fixed by equation \eqref{gpequation}, is obviously different from the one obtained in \cite{Canfora:2013zza}.

The pole structure can be read off from
\begin{eqnarray}
P(p^2)&=&(p^4+\gamma^{\ast \, 4})^2+p^6\xi^{2}a^{2}\sin^{2}(\phi) 
\nonumber\\
&=&(p^2+m_1^2)(p^2+m_2^2)(p^2+m_3^2)(p^2+m_4^2) 
\,.
\label{polynomialequation}
\end{eqnarray}

The first observation about the poles is that it is clearly anisotropic, since it explicitly depends on the azimuthal angle $\phi$. This is a clear consequence of the Lorentz symmetry breaking. Furthermore, despite the considerably nontrivial structure of the poles, as functions of $\xi$, the coupling constant $g$ and the renormalization mass parameter $\bar{\mu}$ (remind that the Gribov parameter $\gamma^{\ast \, 2}$ is a function of the others parameters through the gap equation), we found that they will always be complex conjugated. This is a direct consequence of our perturbative approach with respect to the Lorentz symmetry breaking parameter $\xi$. Therefore, by taking into account the smallness of the upper bound of the CFJ parameter $\xi$, we found that the gauge field propagator cannot describe a physical asymptotic particle. Thus, in Gribov's interpretation, one can say that in this scenario the gauge field is confined.

\section{Conclusion}

In this paper the existence of zero modes of the Faddeev-Popov operator in the Landau gauge, known as the Gribov ambiguity issue, was taken into account in a four dimensional $SU(2)$ Yang-Mills theory with an effective breaking of the Lorentz symmetry implemented by means of the CFJ term. By following Gribov's original approach, we derived the self consistent equation that dynamically determines the Gribov parameter $\beta$, called the Gribov gap equation. We find a strongly modified gap equation, with an explicit anisotropy given by the highly non-trivial integral on the azimuthal angle $\phi$. 

Given the nontrivial gap equation structure in \eqref{gpequation}, a perturbative method was applied, based on the most recent
data for the Lorentz symmetry breaking parameter $\xi=1.1 \times 10^{-7}$ GeV \cite{Kostelecky:2008ts,Colladay:2018jic}, to assess the influence of the CFJ term on the Gribov approach. For example, in a first-order approximation, we were able to find the gap equation \eqref{lehgig34}, with the help of computational efforts. As a matter of qualitative analysis, we defined dimensionless parameters such as $\bar{\beta}^{\frac12} = \beta^{\frac12}/\overline{\mu}^{2}$ and $\bar{\xi} = \xi/\overline{\mu}$, and considered only up to quadratic terms in the Taylor expansion. Besides, we considered $\bar{\xi}$ assuming values around $10^{-9}$ and the coupling constant $g$ assuming random values smaller then $1$. In such a configuration, we found that the Gribov parameter becomes complex valued, although with a considerable small imaginary component (of the order of $10^{-19}$). The real component of the Gribov parameter coincides with the usual (Lorentz symmetric) $SU(2)$ Gribov parameter $\bar{\beta}_{0}^{\frac12}$ with an accuracy of $10^{-2}$. 

Subsequently, we have investigated the behavior of the poles of the gauge field propagator, in accordance with the Gribov parameter numerically derived. Thus, we plugged the values of the Gribov parameter into the poles of the propagators and checked that they will always include an imaginary part. Therefore, we have verified that, by considering the most recent observational bounds to the amplitude of the Lorentz symmetry breaking in a pure non-Abelian gauge theory, the gauge field propagator has a non-physical asymptotic particle interpretation. Thus, it can be interpreted as being confined, according to Gribov's approach.

\section*{Acknowledgments}

We thank to D. Bazeia, C. M. Reyes, F. Canfora, and D. Vercauteren for the fruitful discussions.
I.~F.~J. has been supported by FONDECYT grant
No. $3170278$ and is thankful to the Universidade Federal do Esp\'irito Santo, Brazil, for the
hospitality provided during part of the development of this work. The work by A. Yu. P. has
been partially supported by the CNPq project No. 303783/2015-0.


\begin{thebibliography}{9}

\bibitem{Carroll:1989vb}
  S.~M.~Carroll, G.~B.~Field and R.~Jackiw,
  Phys.\ Rev.\ D {\bf 41} (1990) 1231.



\bibitem{Colladay:1996iz}
  D.~Colladay and V.~A.~Kostelecky,
  Phys.\ Rev.\ D {\bf 55} (1997) 6760
  [hep-ph/9703464].



\bibitem{Colladay:1998fq}
  D.~Colladay and V.~A.~Kostelecky,
  Phys.\ Rev.\ D {\bf 58} (1998) 116002
  [hep-ph/9809521].



\bibitem{Jackiw:2001dj}
  R.~Jackiw,
  Nucl.\ Phys.\ Proc.\ Suppl.\  {\bf 108} (2002) 30
   [Phys.\ Part.\ Nucl.\  {\bf 33} (2002) S6]
   [Lect.\ Notes Phys.\  {\bf 616} (2003) 294]
  [hep-th/0110057].



\bibitem{Guralnik:2001ax}
  Z.~Guralnik, R.~Jackiw, S.~Y.~Pi and A.~P.~Polychronakos,
  Phys.\ Lett.\ B {\bf 517} (2001) 450
  [hep-th/0106044].



\bibitem{Myers:2003fd}
  R.~C.~Myers and M.~Pospelov,
  Phys.\ Rev.\ Lett.\  {\bf 90} (2003) 211601
  [hep-ph/0301124].



\bibitem{Casana:2009xs}
  R.~Casana, M.~M.~Ferreira, Jr, A.~R.~Gomes and P.~R.~D.~Pinheiro,
  Phys.\ Rev.\ D {\bf 80} (2009) 125040
  [arXiv:0909.0544 [hep-th]].



\bibitem{Jackiw:1999qq}
  R.~Jackiw,
  Int.\ J.\ Mod.\ Phys.\ B {\bf 14} (2000) 2011
  [hep-th/9903044].



\bibitem{Kostelecky:2008ts}
  V.~A.~Kostelecky and N.~Russell,
  Rev.\ Mod.\ Phys.\  {\bf 83} (2011) 11
  [arXiv:0801.0287 [hep-ph]].



\bibitem{Santos:2016bqc}
  T.~R.~S.~Santos and R.~F.~Sobreiro,
  Phys.\ Rev.\ D {\bf 94} (2016)  125020
  [arXiv:1607.07413 [hep-th]].



\bibitem{Gomes:2007rv}
  M.~Gomes, J.~R.~Nascimento, E.~Passos, A.~Y.~Petrov and A.~J.~da Silva,
  Phys.\ Rev.\ D {\bf 76} (2007) 047701
  [arXiv:0704.1104 [hep-th]].



\bibitem{Mariz:2007gf}
  T.~Mariz, J.~R.~Nascimento, A.~Y.~Petrov, L.~Y.~Santos and A.~J.~da Silva,
  Phys.\ Lett.\ B {\bf 661} (2008) 312
  [arXiv:0708.3348 [hep-th]].



\bibitem{Santos:2016dcw}
  T.~R.~S.~Santos and R.~F.~Sobreiro,
  Eur.\ Phys.\ J.\ C {\bf 77} (2017)  903
  [arXiv:1612.05538 [hep-th]].



\bibitem{Santos:2016uds}
  T.~R.~S.~Santos, R.~F.~Sobreiro and A.~A.~Tomaz,
  Phys.\ Rev.\ D {\bf 94} (2016)  085027
  [arXiv:1607.05261 [hep-th]].



\bibitem{Santos:2014lfa}
  T.~R.~S.~Santos and R.~F.~Sobreiro,
  Phys.\ Rev.\ D {\bf 91} (2015)  025008
  [arXiv:1404.4846 [hep-th]].



\bibitem{Colladay:2006rk}
  D.~Colladay and P.~McDonald,
  Phys.\ Rev.\ D {\bf 75} (2007) 105002
  [hep-ph/0609084].



\bibitem{Gribov:1977wm}
  V.~N.~Gribov,
  Nucl.\ Phys.\ B {\bf 139} (1978) 1.



\bibitem{Singer:1978dk}
  I.~M.~Singer,
  Commun.\ Math.\ Phys.\  {\bf 60} (1978) 7.



\bibitem{Capri:2015ixa}
  M.~A.~L.~Capri {\it et al.},
  Phys.\ Rev.\ D {\bf 92} (2015) 045039
  [arXiv:1506.06995 [hep-th]].



\bibitem{Capri:2015nzw}
  M.~A.~L.~Capri {\it et al.},
  Phys.\ Rev.\ D {\bf 93} (2016)  065019
  [arXiv:1512.05833 [hep-th]].



\bibitem{Capri:2016aqq}
  M.~A.~L.~Capri {\it et al.},
  Phys.\ Rev.\ D {\bf 94} (2016)  025035
  [arXiv:1605.02610 [hep-th]].



\bibitem{Capri:2016aif}
  M.~A.~L.~Capri, D.~Fiorentini, A.~D.~Pereira, R.~F.~Sobreiro, S.~P.~Sorella and R.~C.~Terin,
  Annals Phys.\  {\bf 376} (2017) 40
  [arXiv:1607.07912 [hep-th]].



\bibitem{Capri:2016gut}
  M.~A.~L.~Capri, D.~Dudal, A.~D.~Pereira, D.~Fiorentini, M.~S.~Guimaraes, B.~W.~Mintz, L.~F.~Palhares and S.~P.~Sorella,
  Phys.\ Rev.\ D {\bf 95} (2017) 045011
  [arXiv:1611.10077 [hep-th]].



\bibitem{Capri:2017bfd}
  M.~A.~L.~Capri, D.~Fiorentini, A.~D.~Pereira and S.~P.~Sorella,
  Phys.\ Rev.\ D {\bf 96} (2017)  054022
  [arXiv:1708.01543 [hep-th]].



\bibitem{Sorella:2009vt}
  S.~P.~Sorella,
  Phys.\ Rev.\ D {\bf 80} (2009) 025013
  [arXiv:0905.1010 [hep-th]].



\bibitem{Dudal:2009bf}
  D.~Dudal, N.~Vandersickel, H.~Verschelde and S.~P.~Sorella,
  PoS QCD {\bf -TNT09} (2009) 012
  [arXiv:0911.0082 [hep-th]].



\bibitem{Capri:2010hb}
  M.~A.~L.~Capri, A.~J.~Gomez, M.~S.~Guimaraes, V.~E.~R.~Lemes, S.~P.~Sorella and D.~G.~Tedesco,
  Phys.\ Rev.\ D {\bf 82} (2010) 105019
  [arXiv:1009.4135 [hep-th]].



\bibitem{Dudal:2012sb}
  D.~Dudal and S.~P.~Sorella,
  Phys.\ Rev.\ D {\bf 86} (2012) 045005
  [arXiv:1205.3934 [hep-th]].



\bibitem{Lavrov:2011wb}
  P.~Lavrov, O.~Lechtenfeld and A.~Reshetnyak,
  JHEP {\bf 1110} (2011) 043
  [arXiv:1108.4820 [hep-th]].



\bibitem{Lavrov:2012gb}
  P.~M.~Lavrov, O.~V.~Radchenko and A.~A.~Reshetnyak,
  Mod.\ Phys.\ Lett.\ A {\bf 27} (2012) 1250067
  [arXiv:1201.4720 [hep-th]].



\bibitem{Sobreiro:2005ec}
  R.~F.~Sobreiro and S.~P.~Sorella,
  hep-th/0504095.



\bibitem{Vandersickel:2012tz}
  N.~Vandersickel and D.~Zwanziger,
  Phys.\ Rept.\  {\bf 520} (2012) 175
  [arXiv:1202.1491 [hep-th]].



\bibitem{Hayashi:2018giz}
  Y.~Hayashi and K.~I.~Kondo,
  Phys.\ Rev.\ D {\bf 99} (2019)  074001
  [arXiv:1812.03116 [hep-th]].



\bibitem{Lavrov:2013boa}
  P.~M.~Lavrov and O.~Lechtenfeld,
  Phys.\ Lett.\ B {\bf 725} (2013) 386
  [arXiv:1305.2931 [hep-th]].



\bibitem{Osterwalder:1973dx}
  K.~Osterwalder and R.~Schrader,
  Commun.\ Math.\ Phys.\  {\bf 31} (1973) 83.



\bibitem{Cucchieri:2007rg}
  A.~Cucchieri and T.~Mendes,
  Phys.\ Rev.\ Lett.\  {\bf 100} (2008) 241601
  [arXiv:0712.3517 [hep-lat]].



\bibitem{Cucchieri:2009zt}
  A.~Cucchieri and T.~Mendes,
  Phys.\ Rev.\ D {\bf 81} (2010) 016005
  [arXiv:0904.4033 [hep-lat]].



\bibitem{Cucchieri:2010xr}
  A.~Cucchieri and T.~Mendes,
  PoS QCD {\bf -TNT09} (2009) 026
  [arXiv:1001.2584 [hep-lat]].



\bibitem{Dudal:2008sp}
  D.~Dudal, J.~A.~Gracey, S.~P.~Sorella, N.~Vandersickel and H.~Verschelde,
  Phys.\ Rev.\ D {\bf 78} (2008) 065047
  [arXiv:0806.4348 [hep-th]].



\bibitem{Dudal:2010tf}
  D.~Dudal, O.~Oliveira and N.~Vandersickel,
  Phys.\ Rev.\ D {\bf 81} (2010) 074505
  [arXiv:1002.2374 [hep-lat]].



\bibitem{Cucchieri:2011ig}
  A.~Cucchieri, D.~Dudal, T.~Mendes and N.~Vandersickel,
  Phys.\ Rev.\ D {\bf 85} (2012) 094513
  [arXiv:1111.2327 [hep-lat]].



\bibitem{Cucchieri:2012cb}
  A.~Cucchieri, D.~Dudal and N.~Vandersickel,
  Phys.\ Rev.\ D {\bf 85} (2012) 085025
  [arXiv:1202.1912 [hep-th]].



\bibitem{Cucchieri:2013nja}
  A.~Cucchieri and T.~Mendes,
  Phys.\ Rev.\ D {\bf 88} (2013) 114501
  [arXiv:1308.1283 [hep-lat]].






\bibitem{Alkofer:2000wg}
  R.~Alkofer and L.~von Smekal,
  Phys.\ Rept.\  {\bf 353} (2001) 281
  [hep-ph/0007355].





\bibitem{Watson:2001yv}
  P.~Watson and R.~Alkofer,
  Phys.\ Rev.\ Lett.\  {\bf 86} (2001) 5239
  [hep-ph/0102332].




\bibitem{Alkofer:2003jj}
  R.~Alkofer, W.~Detmold, C.~S.~Fischer and P.~Maris,
  Phys.\ Rev.\ D {\bf 70} (2004) 014014
  [hep-ph/0309077].





\bibitem{Aguilar:2004sw}
  A.~C.~Aguilar and A.~A.~Natale,
  JHEP {\bf 0408} (2004) 057
  [hep-ph/0408254].



\bibitem{Aguilar:2006gr}
  A.~C.~Aguilar and J.~Papavassiliou,
  JHEP {\bf 0612} (2006) 012
  [hep-ph/0610040].



\bibitem{Aguilar:2008xm}
  A.~C.~Aguilar, D.~Binosi and J.~Papavassiliou,
  Phys.\ Rev.\ D {\bf 78} (2008) 025010
  [arXiv:0802.1870 [hep-ph]].





\bibitem{Strauss:2012dg}
  S.~Strauss, C.~S.~Fischer and C.~Kellermann,
  Phys.\ Rev.\ Lett.\  {\bf 109} (2012) 252001
  [arXiv:1208.6239 [hep-ph]].






\bibitem{Maas:2011se}
  A.~Maas,
  Phys.\ Rept.\  {\bf 524} (2013) 203
  [arXiv:1106.3942 [hep-ph]].





\bibitem{Huber:2018ned}
  M.~Q.~Huber,
  arXiv:1808.05227 [hep-ph].




\bibitem{Tissier:2010ts}
  M.~Tissier and N.~Wschebor,
  Phys.\ Rev.\ D {\bf 82} (2010) 101701
  [arXiv:1004.1607 [hep-ph]].



\bibitem{Tissier:2011ey}
  M.~Tissier and N.~Wschebor,
  Phys.\ Rev.\ D {\bf 84} (2011) 045018
  [arXiv:1105.2475 [hep-th]].



\bibitem{Pelaez:2014mxa}
  M.~Pelaez, M.~Tissier and N.~Wschebor,
  Phys.\ Rev.\ D {\bf 90} (2014) 065031
  [arXiv:1407.2005 [hep-th]].



\bibitem{Zwanziger:1993dh}
  D.~Zwanziger,
  Nucl.\ Phys.\ B {\bf 412} (1994) 657.



\bibitem{Cucchieri:1997ns}
  A.~Cucchieri,
  Nucl.\ Phys.\ B {\bf 521} (1998) 365
  [hep-lat/9711024].



\bibitem{Cucchieri:2007md}
  A.~Cucchieri and T.~Mendes,
  PoS LATTICE {\bf 2007} (2007) 297
  [arXiv:0710.0412 [hep-lat]].



\bibitem{Cucchieri:2008fc}
  A.~Cucchieri and T.~Mendes,
  Phys.\ Rev.\ D {\bf 78} (2008) 094503
  [arXiv:0804.2371 [hep-lat]].



\bibitem{Bogolubsky:2009qb}
  I.~L.~Bogolubsky, E.-M.~Ilgenfritz, M.~Muller-Preussker and A.~Sternbeck,
  PoS LAT {\bf 2009} (2009) 237
  [arXiv:0912.2249 [hep-lat]].



\bibitem{Dudal:2009xh}
  D.~Dudal, S.~P.~Sorella, N.~Vandersickel and H.~Verschelde,
  Phys.\ Rev.\ D {\bf 79} (2009) 121701
  [arXiv:0904.0641 [hep-th]].



\bibitem{Sorella:2010it}
  S.~P.~Sorella,
  J.\ Phys.\ A {\bf 44} (2011) 135403
  [arXiv:1006.4500 [hep-th]].



\bibitem{Jackiw:1999yp}
  R.~Jackiw and V.~A.~Kostelecky,
  Phys.\ Rev.\ Lett.\  {\bf 82} (1999) 3572
  [hep-ph/9901358].



\bibitem{Mariz:2005jh}
  T.~Mariz, J.~R.~Nascimento, E.~Passos, R.~F.~Ribeiro and F.~A.~Brito,
  JHEP {\bf 0510} (2005) 019
  [hep-th/0509008].




\bibitem{Dudal:2017jfw}
  D.~Dudal and D.~Vercauteren,
  Phys.\ Lett.\ B {\bf 779} (2018) 275
  [arXiv:1711.10142 [hep-th]].



\bibitem{Kroff:2018ncl}
  D.~Kroff and U.~Reinosa,
  Phys.\ Rev.\ D {\bf 98} (2018)  034029
  [arXiv:1803.10188 [hep-th]].



\bibitem{Colladay:2018jic}
  D.~Colladay, J.~P.~Noordmans and R.~Potting,
  J.\ Phys.\ Conf.\ Ser.\  {\bf 952} (2018)  012021.



\bibitem{Canfora:2013zza}
  F.~Canfora, A.~J.~Gomez, S.~P.~Sorella and D.~Vercauteren,
  Annals Phys.\  {\bf 345} (2014) 166
  [arXiv:1312.3308 [hep-th]].








  
  
  

\end{thebibliography}
\end{document}